\documentclass{aa}
\usepackage{graphicx}
\usepackage[sort]{natbib}
\bibpunct{(}{)}{;}{a}{}{,}
\DeclareTextSymbol{\degre}{OT1}{23}

\def \darwin {D{\sc arwin}}
\def \peg {P{\sc egase}}

\usepackage{txfonts}

\begin{document}
   \title{Nulling interferometry: performance comparison between space and ground-based sites for exozodiacal disc detection}
   \titlerunning{Nulling interferometry: performance comparison between ground and space for exozodiacal disc detection}

   \author{D. Defr\`ere\inst{1}
          \and
          O. Absil\inst{2}\fnmsep\thanks{Marie Curie EIF Postdoctoral Fellow.} 
          \and
          V. Coud\'e du Foresto\inst{3}
          \and
          W.C. Danchi\inst{4}
          \and
          R. den Hartog\inst{5}
          }

   \offprints{D. Defr\`ere}

   \institute{Institut d'Astrophysique et de G\'eophysique, Universit\'e de Li\`ege, 17 All\'ee du Six
 Ao\^ut, B-4000 Li\`ege, Belgium\\
              \email{defrere@astro.ulg.ac.be}
         \and
             LAOG--UMR 5571, CNRS and Universit\'e Joseph Fourier, BP 53, F-38041 Grenoble, France
         \and
             LESIA, Observatoire de Paris-Meudon, CNRS, 5 place
             Jules Janssen, F-92195 Meudon, France
         \and
             NASA Goddard Space Flight Center, 8800, Greenbelt
             Road, Greenbelt, MD 20771, USA
         \and
             Science Payloads and Advanced Concepts Office, ESA/ESTEC, postbus 299, NL-2200 AG Noordwijk, The Netherlands
             }

   \date{Received 23 May 2008; accepted 11 August 2008}

  \abstract
   {Characterising the circumstellar dust around nearby main sequence stars is
    a necessary step in understanding the planetary formation
    process and is crucial for future life-finding
    space missions such as ESA's \darwin{} or NASA's Terrestrial
    Planet Finder (TPF). Besides paving the technological
    way to \darwin/TPF, the space-based infrared interferometers
    \peg{} and FKSI (Fourier-Kelvin Stellar Interferometer) will be valuable scientific precursors.}
   {We investigate the performance of \peg{} and FKSI for exozodiacal disc detection and
    compare the results with ground-based nulling interferometers.}
   {We used the GENIEsim software (Absil et al. 2006) which was designed and validated to study the performance of ground-based nulling interferometers.
   The software has been adapted to simulate the performance of space-based nulling interferometers by disabling all atmospheric effects and
   by thoroughly implementing the perturbations induced by payload vibrations in the ambient space environment.}
   {Despite using relatively small telescopes ($\leq$ 0.5\,m), \peg{} and FKSI are very efficient for exozodiacal disc
    detection. They are capable of detecting exozodiacal
    discs respectively 5 and 1 time as dense as the solar zodiacal
    cloud and they outperform any ground-based instrument. Unlike \peg, FKSI can achieve this sensitivity for most targets of
    the \darwin/TPF catalogue thanks to an appropriate combination of baseline length
    and observing wavelength. The sensitivity of \peg{} could, however, be significantly boosted by
    considering a shorter interferometric baseline length.}
   {Besides their main scientific goal (characterising hot giant extrasolar planets), the space-based nulling
   interferometers \peg{} and FKSI will be very efficient in assessing within a few minutes the level of circumstellar dust
   in the habitable zone around nearby main sequence stars down to the density of the solar zodiacal
   cloud. These space-based interferometers would be complementary
   to Antarctica-based instruments in terms of sky coverage and would be ideal instruments
   for preparing future life-finding space missions.}

   \keywords{Instrumentation: high angular resolution --
             techniques: interferometric --
             circumstellar matter
               }

   \maketitle
%

\section{Introduction}

\begin{figure*}[t]
\centering
\includegraphics[width=7cm]{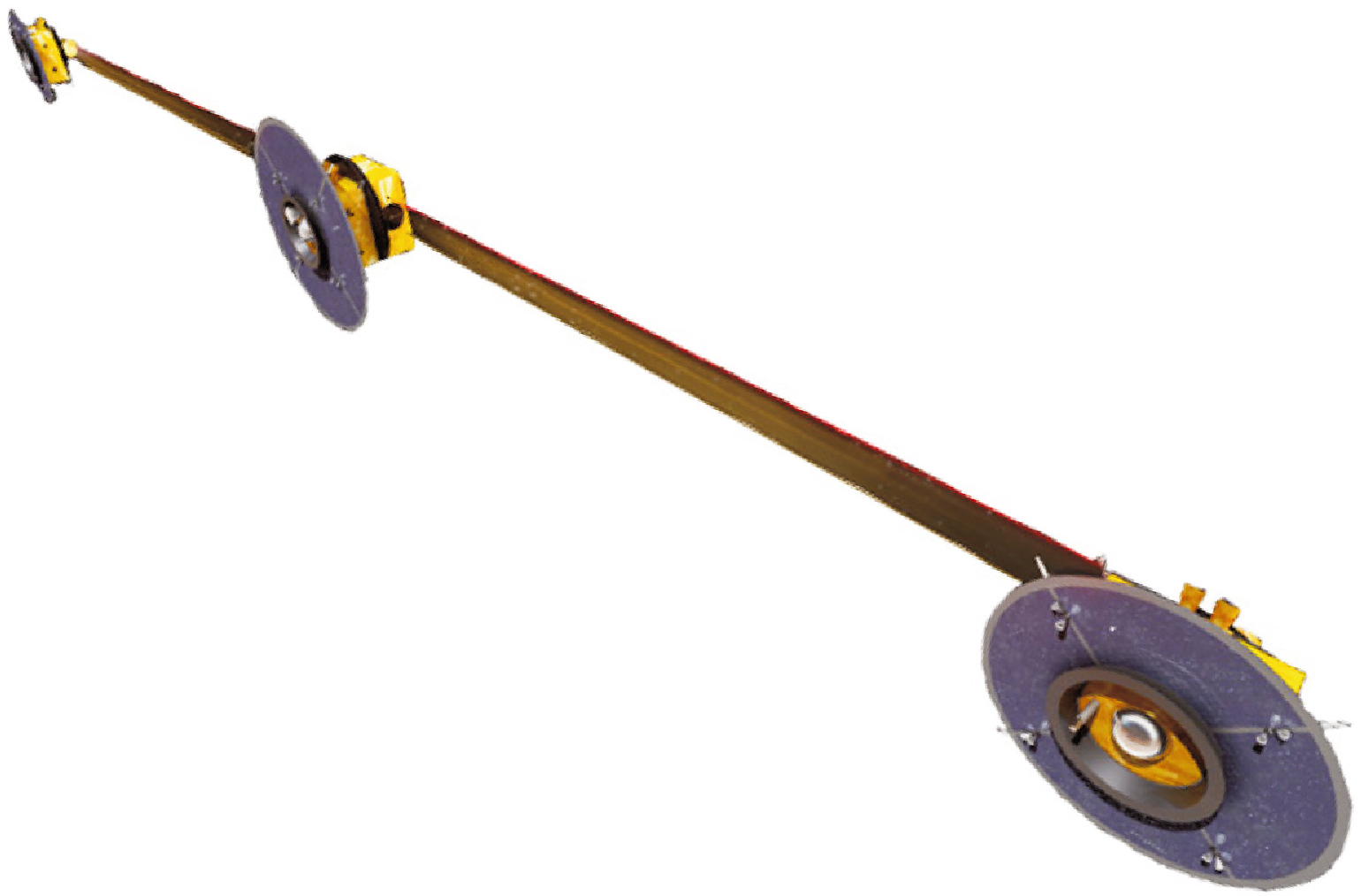}\hspace{1.5 cm}
\includegraphics[width=6cm]{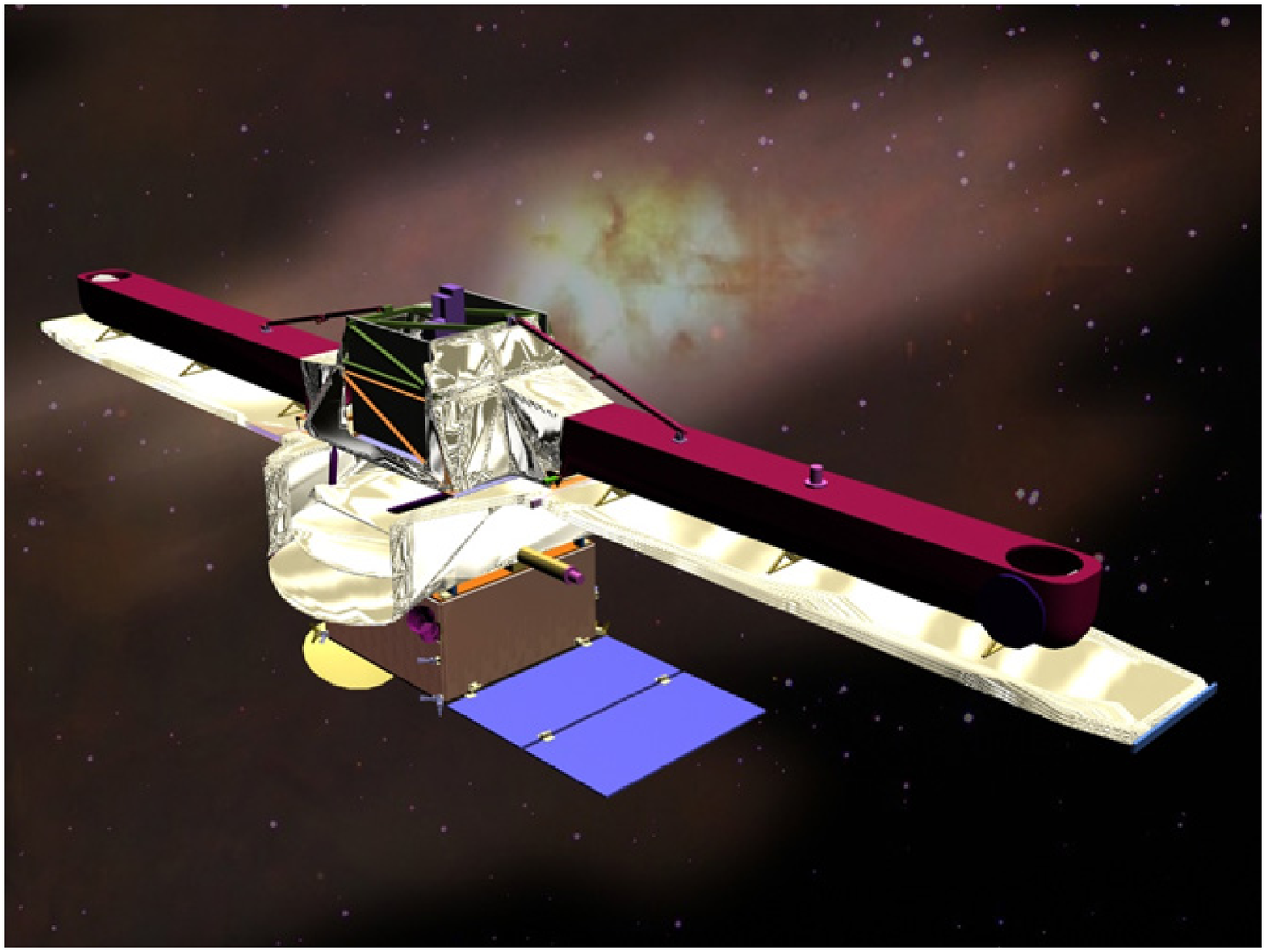}
\caption{Left: overview of the \peg{} space-based interferometer.
Two 0.4-m siderostats are flying in a linear configuration with
the beam combiner spacecraft located in the middle of the
formation. Right: representation of FKSI, showing the two 0.5-m
siderostats located on a 12.5-m boom.} \label{fig:pegase_FKSI}
\end{figure*}

Nulling interferometry is the core technique of future
life-finding space missions such as ESA's \darwin{}
\citep{Fridlund:2006} and NASA's Terrestrial Planet Finder
Interferometer \citep[TPF-I,][]{Beichman:2006}. Observing in the
mid-infrared (6-20\,$\mu$m), these missions would enable the
spectroscopic characterisation of the atmosphere of habitable
extrasolar planets orbiting nearby main sequence stars. This
ability to study habitable distant planets strongly depends on the
density of exozodiacal dust in the inner part of circumstellar
discs, where the planets are supposed to be located. In
particular, the detection of habitable terrestrial planets would
be seriously hampered for stars presenting warm ($\sim$300\,K)
exozodiacal dust more than 10 to 100 times as dense as our solar
zodiacal disc, depending on stellar type, stellar distance and
telescope diameter \citep{beichman:2006b,defrere:2008}. Assessing
the level of circumstellar dust around nearby main sequence stars
is therefore a necessary pre-requisite for preparing the observing
programme of \darwin/TPF by reducing the risk of wasting time on
sources for which exozodiacal light prevents Earth-like planet
detection. In addition, the existence of planets is intrinsically
linked to circumstellar discs and observing them provides an
efficient way to study the formation, evolution and dynamics of
planetary systems. At young ages, essentially all stars are
surrounded by protoplanetary discs in which the planetary systems
are believed to form \citep{Meyer:2008}. In particular, the
detection of gaps in these protoplanetary discs is very important
for understanding the early dynamics of planets, including migration
and orbital interaction. At older ages, photometric surveys
primarily with IRAS, ISO, and Spitzer have revealed the presence
of micron-sized grains around a large number of main sequence
stars \citep[see e.g.,][]{Trilling:2008,Hillebrand:2008}. This is
interpreted as the sign of planetary activity, as the production
of grains is believed (by analogy with the zodiacal cloud in our
solar system) to be sustained by asteroid collisions and
outgassing of comets in the first tens of astronomical units (AU).
However, the presence of warm dust can generally not be
unequivocally determined because the typical accuracy on both
near-infrared photometric measurements and photospheric flux
estimations is a few percent at best, limiting the sensitivity to
typically 1000 times the density of our solar zodiacal cloud
\citep{Beichman:2006c}. Photometric measurements are therefore
generally not sufficient to probe the innermost regions of the
discs and interferometry is required to separate the starlight
from the disc emission. Good examples are given by the detection
of hot dust ($\sim$1500\,K) around Vega and $\tau$ Cet
with near-infrared interferometry at the CHARA array
\citep{Absil:2006b,Difolco:2007}. Nulling interferometry is a
quite new technique even though it was initially proposed in 1978
\citep{Bracewell:1978}. Several scientific observations using this technique
have recently been carried out with the Bracewell Infrared Nulling
Cryostat (BLINC, \citealt{Hinz:2000}) instrument at the
Multi-Mirror Telescope (MMT, Mont Hopkins, Arizona), with the Keck
Interferometer Nuller (KIN, Hawaii,
\citealt{Serabyn:2006,Barry:2008,Serabyn:2008}), and are foreseen to
begin in 2010 at the Large Binocular Telescope (Mount Graham,
Arizona,\citealt{Hinz:2008}). In Europe, ESA has initiated the study
of a ground-based demonstrator for \darwin, the Ground-based European
Nulling Interferometer Experiment \citep[GENIE,][]{gondoin:2004}. 
GENIE is a nulling interferometer conceived as a focal instrument for the VLTI which has been
studied by ESA at the phase A level. Another European project is
ALADDIN (Antarctic L-band Astrophysics Discovery Demonstrator for
Interferometric Nulling, \citealt{Foresto:2006}), a nulling
interferometer project for Dome C, on the high Antarctic plateau.
The performance of GENIE has been studied in detail
(\citealt{Absil:2006_paper1},\citealt{Wallner:2006}) and recently
compared to that of ALADDIN \citep{Absil:2007_paper2}. Using 1-m
collectors, ALADDIN would have an improved sensitivity with
respect to GENIE working on 8-m telescopes, provided that it is
placed above the turbulence boundary layer (about 30\,m at Dome
C). Circumstellar discs 30 times as dense as our local zodiacal
cloud could be detected by ALADDIN around typical \darwin/TPF
targets in an integration time of few hours.

The low atmospheric turbulence on the high Antarctic plateau is a
significant advantage with respect to other astronomical sites and
one of the main reasons for the very good sensitivity of ALADDIN.
However, as for any other ground-based site, the atmosphere effects (turbulence and thermal background) 
are still major limitations to the
performance and active compensation by real-time control systems
are mandatory. Observing from space would provide an efficient
solution to improve the sensitivity by getting rid of the harmful
effect of the atmosphere. Two infrared nulling interferometers
could achieve the detection of circumstellar dust discs from space
(see Fig.~\ref{fig:pegase_FKSI}): \peg, a two-telescope
interferometer based on three free-flying spacecraft
\citep{Leduigou:2006} and the Fourier-Kelvin Stellar
interferometer (FKSI), a structurally-connected interferometer
also composed of two telescopes \citep{Danchi:2006}. These two
missions have been initially designed to study hot extrasolar
giant planets at high angular resolution in the near- to mid-infrared
regime (respectively 1.5-6.0 $\mu$m and 3.0-8.0 $\mu$m). Besides
their main scientific goal, they could also be particularly well
suited for the detection of warm circumstellar dust in the
habitable zone around nearby main sequence stars. The objective
would be to provide a statistically significant survey of the
amount of exozodiacal light in the habitable zone around the
\darwin/TPF targets, and its prevalence as a function of other
stellar characteristics (age, spectral type, metallicity, presence
of a cold debris disc, etc). Following our performance studies of
ground-based instruments such as GENIE at Cerro Paranal
(\citealt{Absil:2006_paper1}, hereafter Paper\,I) or ALADDIN on
the high Antarctic plateau (\citealt{Absil:2007_paper2}, hereafter
Paper\,II), the present study addresses the performance of
space-based nulling instruments for exozodiacal disc detection.
We have limited our comparison to instruments working at
similar wavelengths (ranging from 2 to 8 $\mu$m), and purposely
discarded ground-based instruments working in the N-band such as
the KIN and the LBTI. The ultimate performance of these two
mid-infrared instruments essentially depends on the spatial and
temporal fluctuations of the sky and instrumental thermal
backgrounds, which are very difficult to model with a sufficient
accuracy for our comparative study.


\section{P{\small EGASE} and FKSI overview}

\peg{} and FKSI are space-based Bracewell interferometers,
conceived as scientific and technological precursors to
\darwin/TPF. They present similar architectures, the main
difference being that the two telescopes of \peg{} are free-flying
while those of FKSI are arranged on a single boom. \peg{} was
initially proposed in the framework of the 2004 call for ideas by
the French space agency (CNES) for its formation flying
demonstrator mission. CNES performed a Phase 0 study in 2005 and
concluded that the mission is feasible within an 8 to 9 years
development plan \citep{Leduigou:2006}. However, the mission was
not selected for budgetary reasons. On the US side, FKSI has been
initially studied by the Goddard Space Flight Center in
preparation for submission as a Discovery-class mission. Several
concepts have been considered and the mission was studied to the
phase A level based on the two-telescope design described here.

\subsection{Scientific objectives}

The main scientific goal of \peg{} and FKSI is to perform the
spectroscopy of hot extrasolar giant planets (EGP). With a minimum
baseline length of 40\,m, \peg{} could directly survey most hot
Jupiter-like planets (M $\geq$ 0.2 M$_{\rm Jup}$) within 150\,pc
with a good signal-to-noise ratio (SNR) as well as several
favourable hot Uranus-type planets (0.04 M$_{\rm Jup}$ $\leq$ M
$\leq$ 0.2 M$_{\rm Jup}$) with a final SNR ranging between 1 and 6
\citep{Absil:thesis}. In particular, \peg{} will be able to
perform spectroscopy on about 15\% of the extrasolar planets known
so far within 25\,pc, including several planets outside the hot
regime (further than 0.1 AU from the host star,
\citealt{Defrere:2007}). The working method of FKSI is slightly
different from that of \peg. Due to its relatively short baseline
length (12.5\,m), FKSI uses a two-color method (based on the ratio
of measurements at two wavelengths) to account for the fact that
the planetary signal is likely to fall partly within the central
dark fringe \citep{Danchi:2003b}. Using this method, an
earlier version of FKSI was estimated to be able to detect at least 25 EGPs,
obtain low resolution spectra of their atmosphere and make precise
determination of their orbital parameters \citep{Barry:2006}. This
previous version of FKSI presented an 8-m boom, assumed
15\,nm rms residual OPD error and considered a sample of 140 known
extrasolar planets. With the current version of FKSI, as discussed
in this paper (12.5-m boom length and 2-nm rms residual OPD errors), and
considering a much larger available sample of known extrasolar
planets ($\sim$250), this value should be of the order of 75-100.
Work is in progress to determine how many known extrasolar planets
can be detected with FKSI, as well as the possibility of detecting
super-Earths.

The detection and characterisation of circumstellar discs are also
in the core programmes of these two missions but the performance
has not yet been carefully assessed. \peg{} and FKSI are expected
to be able to provide an accurate estimate of the dust density
from the very neighbourhood of the star up to several AUs. They
will also help providing maps of the mineralogical composition,
with a combination of spectral and spatial information on the
discs. Combined with sub-mm observations from the ground providing
the gas distribution with a comparable spatial resolution, it will
then become possible to study the dust-gas interactions in young
systems. Additional programmes on brown dwarfs and active galactic
nuclei are also foreseen, but only the primary objective (the
study of hot EGPs) drives the design of the instruments.

\subsection{The P{\small EGASE} instrumental concept}\label{sec:pegase}

\begin{table}[t]
  \centering
  \caption{Instrumental parameters of \peg{} and FKSI considered in this study.}\label{Tab:peg_Spec}
  \begin{tabular}{c}
    \hline
    \hline
    \begin{tabular}{l c c}
      Instrumental parameters & \peg{} & FKSI\\
      \hline
      Baselines [m] & 40-500 & 12.5 \\
      Telescope diameter [m] & 0.40 & 0.50\\
      Field of regard & $\pm$ 30\degre & $\pm$ 20\degre \\
      Optics temperature [K] & 90 & 65  \\
      Detector temperature [K] & 55 & 35  \\
      Science waveband [$\mu$m] & 1.5-6.0 & 3.0-8.0 \\
      Spectral resolution & 60 & 20\\
      Fringe sensing waveband [$\mu$m] & 0.8-1.5 & 0.8-2.5 (80\%)\\
      Tip-tilt sensing waveband [$\mu$m] & 0.6-0.8 & 0.8-2.5 (20\%)\\
    \end{tabular}\\
    \hline
  \end{tabular}  
\end{table}

\begin{figure}[t]
\begin{flushleft}
\hspace{-0.3cm}
\includegraphics[width=8cm]{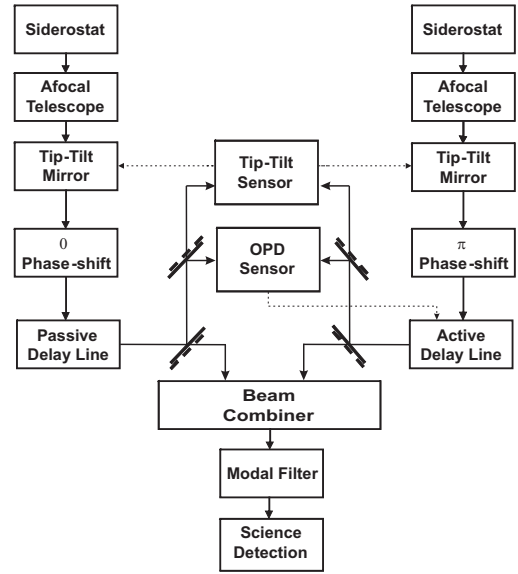}
\caption{Block diagram of the \peg/FKSI optical layout. Feed-back
signals driving the tip-tilt/OPD control are represented by dashed
lines.} \label{fig:block_diag} \end{flushleft}
\end{figure}

Following the phase 0 study, the baseline configuration of \peg{}
consists in a two-aperture near-infrared (1.5-6 $\mu$m)
interferometer formed of three free flying spacecraft planned to
orbit at the Lagrange point L2, where the spacecraft and the focal plane assembly can be passively
cooled down to respectively 90\,K and 55\,K. In its nominal
configuration, \peg{} consists in two 40 cm siderostats and a beam
combiner flying in linear formation. Visibility measurements and
recombination in nulling mode (Bracewell interferometer) are both
possible with a spectral resolution of about 60. The
interferometric baseline length ranges between 40\,m and 500\,m
giving an angular resolution in the range of 0.5-30 mas. Shorter
baseline lengths are not allowed due to the free-flying collision
avoidance distance of 20\,m. The fine-tuning of the optical path
difference (OPD) is performed by a dedicated control loop based on
a fringe sensing unit (FSU) using the observed central target in
the 0.8-1.5 $\mu$m range and an optical delay line (ODL).
Intensity control is performed by a fine pointing loop using a
field relative angle sensor (FRAS) operating in the 0.6-0.8 $\mu$m
range and fast steering mirrors based on piezoelectric devices.
The instrumental parameters of \peg{} are summarized in
Table~\ref{Tab:peg_Spec}. The optical system architecture is
represented by the block diagram in Fig.~\ref{fig:block_diag} with
the following elements on the optical path:

\begin{itemize}
\item Two afocal telescopes with an optical magnification which
will result from a trade-off between the dynamics of the tip-tilt
errors, the available stroke of the fast steering mirrors, the
actuation noise, the mechanical constraints and the polarization
limitations. A magnification of the order of 20 is considered in
the present design. \item Two fast steering mirrors to correct the
tip-tilt errors. They are placed as close as possible to the
afocal telescopes in order to minimize the optical path where the
tip-tilt errors are not corrected, and hence reduce differential
polarisation effects. \item The achromatic $\pi$ phase-shift is
achieved geometrically, by means of opposite periscopes producing
field reversal by reflections \citep{Serabyn:1999}. \item Two
optical delay lines placed after the active mirrors to operate in
a tip/tilt corrected optical space. \item Dichroic beam splitters
which separate the signal between the science wave band and the
tip-tilt/OPD sensing wave bands.\item A Modified Mach Zehnder
\citep[MMZ,][]{Serabyn:2001} to perform beam combination. A second
MMZ might be necessary to cover the full wavelength range,
depending on the coatings. \item Small off axis parabolas to focus
the four outputs of the MMZ into single mode fibres. A fluoride
glass fibre can cover the spectral range 1.5-3 $\mu$m. A
chalcogenide fibre is required for the spectral range 3-6 $\mu$m.
\item A detection assembly controlled at a temperature of 55\,K
and connected to the fibres.
\end{itemize}

\subsection{The FKSI instrumental concept}

Resulting from several dedicated studies in the past few years,
the FKSI design nowadays consists in two 0.5-m telescopes on a
12.5-m boom. The wavelength band used for science ranges from 3
to 8 $\mu$m, which gives an angular resolution between
about 25 and 66\,mas. The instrument is foreseen to be launched to
L2 where it will be passively cooled down to 65\,K. The field of
regard is somewhat smaller than the one of \peg{} with possible
angles of $\pm$ 20\degre\,around the anti-solar direction (vs
$\pm$ 30\degre\,for \peg). This value depends on the size of the
sunshields considered in the present design and could eventually
be increased. The optical arrangement is similar to that of \peg{}
and follows the description given in Section \ref{sec:pegase},
with some differences explained hereafter (see also
Fig.~\ref{fig:block_diag}). OPD stabilization is performed by a
FSU using the observed central target in the 0.8-2.5 $\mu$m range
and feeding an ODL. Unlike \peg, tip/tilt control is performed in
the same wavelength range as the OPD control. After separation
from the science signal with dichroic beam splitters, 80\% of the
light in the 0.8-2.5 $\mu$m range feeds the FSU and 20\% the
tip/tilt sensor. One hollow-glass fibre is used as modal filter in
the 3.0-8.0 $\mu$m wavelength range at each of the two destructive
outputs of a symmetric Mach Zehnder beam combiner
\citep{Barry:2006}. The fibres outputs are focused on the science
detector, cooled down to a temperature of 35 K. Note that photonic
crystal fibres are also considered and are a promising solution
for single mode propagation on a wider spectral band. The
instrumental parameters of FKSI are listed in
Table~\ref{Tab:peg_Spec}.


%

\section{Nulling performance in space}

In order to assess the performance of \peg{} and FKSI for
exozodiacal disc detection, the GENIE simulation software
(GENIEsim, see Paper\,I) has been used. GENIEsim has originally
been designed to simulate the GENIE instrument at the VLTI
interferometer and has been extensively validated by
cross-checking with performance estimates done by industrial
partners during the GENIE phase A study. GENIEsim performs
end-to-end simulations of ground-based nulling interferometers,
including the simulation of astronomical sources (star,
circumstellar disc, planets, background emission), atmospheric
turbulence (piston, longitudinal dispersion, wavefront errors,
scintillation), as well as a realistic implementation of
closed-loop compensation of atmospheric effects by means of fringe
tracking and wavefront correction systems. The output of the
simulator basically consists in time series of photo-electrons
recorded by the detector at the constructive and destructive
outputs of the nulling combiner. To enable the simulation of a
space-based nulling interferometer, few modifications were
necessary due to the versatility of GENIEsim. Beside disabling all
atmospheric effects, the main modification was to introduce the
random sequences of OPD and tip/tilt generated by the vibrations
of the telescopes in the ambient space environment. This is
discussed in the following section.

\begin{table*}[!t]
\centering \caption{Control loop performance and optimum
repetition frequencies computed on a 100 sec observation sequence for a Sun-like G2V star located at 20\,pc.
The total null is the mean nulling ratio including both the geometric and instrumental
leakage contributions. The rms null is the standard deviation of the instrumental nulling ratio for this 100 sec sequence. The goal performance for
exozodiacal disc detection discussed in Paper\,I appears in the last column.} \label{Fig:loop_perfo}
\begin{tabular}{c c c c c|c}
  \hline
  \hline
    & GENIE-UT & ALADDIN & \peg{} & FKSI & Goal \\
  \hline
  Piston & 6.2 nm @ 13 kHz &  10 nm @ 2 kHz & 1.7 nm @ 60 Hz & 2 nm @ 65 Hz & $<$ 4nm \\
  Inter-band disp. & 4.4 nm @ 300 Hz &  7.0 nm @ 0 kHz & 0 nm @ 0 kHz & 0 nm @ 0 kHz & $<$ 4nm \\
  Intra-band disp. & 1.0 nm @ 300 Hz &  7.4 nm @ 0 Hz & 0 nm @ 0 kHz & 0 nm @ 0 kHz & $<$ 4nm \\
  Tip-tilt & 11 mas @ 1 kHz &  7 mas @ 1 kHz & 15 mas @ 85 Hz & 20 mas @ 60 Hz& (see intensity) \\
  Intensity mismatch & 4\% @ 1 kHz &  1.2\% @ 0 Hz & 0.02\% @ 0 kHz & 0.04\% @ 0 Hz & $<$ 1\% \\
  \hline
  Total null & 6.2 $\times$ 10$^{-4}$ &  2.2 $\times$ 10$^{-4}$ & 1.0 $\times$ 10$^{-3}$ & 3.0 $\times$ 10$^{-5}$ & $f(b,\lambda)$ \\
  Instrumental null & 1.5 $\times$ 10$^{-4}$ &  1.3 $\times$ 10$^{-4}$ & 1.0 $\times$ 10$^{-5}$ & 7.0 $\times$ 10$^{-6}$ & 10$^{-5}$ \\
  RMS null & 2.0 $\times$ 10$^{-6}$ &  3.5 $\times$ 10$^{-6}$ & 1.1 $\times$ 10$^{-7}$ & 6.9 $\times$ 10$^{-8}$ & 10$^{-5}$ \\
  \hline
  \end{tabular}
\end{table*}

\subsection{Vibrations in space environment}

\begin{figure}[!t]
\centering
\hspace{-0.5 cm}
\includegraphics[width=9.2cm]{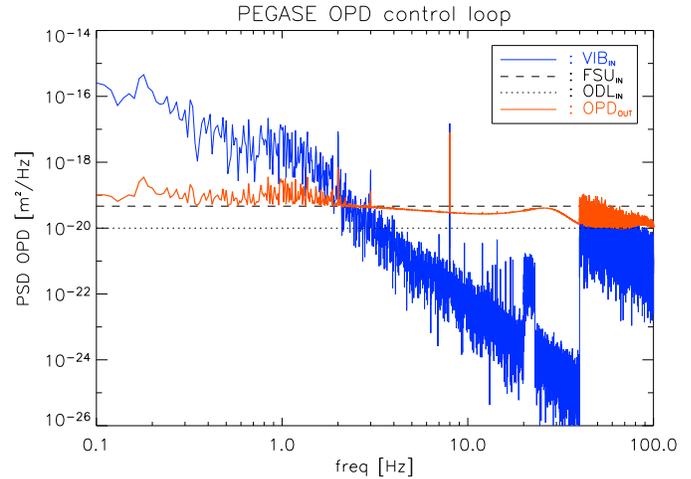}
\caption{Power spectral density of OPD errors at the input (VIB$_{\rm 
IN}$) and at the output (OPD$_{\rm OUT}$) of the \peg{} control
loop. The PSDs of the fringe sensing unit (FSU$_{\rm IN}$, dashed
curve) and of the ODL (ODL$_{\rm IN}$, dotted curve) before the
OPD control loop are also represented. Similar PSDs are used for
FKSI, taking into account in the input OPD perturbations an
additional 5-Hz contribution due to the boom.} \label{fig:psd_opd}
\end{figure}

Spacecraft vibrations are critical in nulling interferometry
because they induce fluctuations in the differential optical paths
and pointing errors, which both give rise to stochastic stellar
leakage in the destructive output. These vibrations are caused by
disturbance forces which can be either internal (due to on-board
systems) or external (caused by the ambient space environment).
Internal disturbance forces arise mainly from the thrusters, the
Optical Delay Line (ODL), the steering mirrors, the reaction
wheels and the boom in the case of structurally connected
telescopes. The external disturbance forces are mainly caused by
particulate impacts, solar radiation pressure and charging effects
but all these effects are not expected to be dominant at the L2
point. A recent comparative study concludes that the ambient space
environment causes OPD errors no larger than the disturbances
induced by on-board equipment for \darwin{} and its precursor
missions \citep{Sterken:2005}.

In the case of \peg, an R\&D study carried out by EADS-Astrium in
collaboration with CNES \citep{Villien:2007} has identified the
two main sources of perturbations: torque noise and
micro-vibrations, both at the reaction wheel level. The torque
noise corresponds to the perturbations around the wheel rotation
axis. It is due to the wheel electronics noise, the wheel
controller loop, the friction torque and the motor defect.
Micro-vibrations are due to the wheel mechanical defects
such as rotor imbalance and ball bearing imperfections. They
correspond to harmonic perturbations, function of the wheel
velocity and generate both torques and forces disturbances. In the
present architecture, the observation is considered to be divided
into a succession of 100-s phases of science and control: pulse
control phases of 100 s interrupt the science observation during
which the constellation is free flying (so that there is no
thruster noise during this phase). The Power Spectral Densities
(PSD) defined by Astrium and CNES during the R\&D study for OPD
and tip/tilt have been implemented in GENIEsim. The PSD of the OPD
in m$^2$/Hz is represented in Fig.~\ref{fig:psd_opd} by the solid
curve labelled ``VIB$_{\rm IN}$" and defines the vibrational level
at the input of the FSU (about 0.18\,$\mu$m rms). This PSD
corresponds to a wheel rotation frequency of 1\,Hz, with flexible
modes at 20\,Hz (due to the sunshield) and 40\,Hz (due to the
platform structure). Increasing the rotation frequency of the
wheel could reduce the torque noise but at the expense of
micro-vibrations. The shape of the tip/tilt PSD is similar to that
of OPD with a value of about $3\farcs5$ rms at the input of the
tip/tilt sensor.

For FKSI, the reaction wheels are also expected to be the main
contributor to the vibrational level \citep{Tupper:2004}. Another
contribution comes from boom deflections induced by thermal
changes and producing low frequency OPD. The worst case occurs at
the boom resonant frequency which results in a sine wave with an
amplitude of 2.4\,nm at 5\,Hz for the OPD perturbation and a sine
wave of 0.2\,mas at 5\,Hz for the tip/tilt perturbation (Tupper
Hyde, private communication). Assuming that FKSI will use the same
wheels as \peg, we can in good approximation use the PSD defined
for \peg, to which we add the resonant boom contribution at 5\,Hz.

\subsection{Control loop performance}

As indicated in the previous section, the level of OPD and
tip/tilt would be of the order of 0.18\,$\mu$m and $3\farcs5$ rms
without appropriate correction techniques. This is prohibitive for
exozodiacal disc detection and fine control loops are therefore
mandatory to stabilise the OPD and the tip-tilt to acceptable values. In
GENIEsim, control loops are simulated through their transfer
function in the frequency domain. A simultaneous optimisation
is performed on the loop repetition frequency and the controller
parameters (a simple PID\footnote{PID stands for ``Proportional,
Integral and Differential" which is a basic controller device for
closed-loop control}) in order to minimise the residual errors, which 
are computed by integrating the corrected PSD on
the frequency domain. The PSDs of the OPD perturbation before and after fringe
tracking are shown in Fig.~\ref{fig:psd_opd} in the case of \peg.
At the input of the loop, the OPD perturbations come from the
wheels (VIB$_{\rm IN}$), the FSU measurement noise (FSU$_{\rm
IN}$) and the intrinsic ODL noise (ODL$_{\rm IN}$). The PSD of the
FSU noise is computed by considering a standard ABCD algorithm to
estimate the phase of the fringe and assuming a read-out noise of
15 electrons rms per pixel. For the ODL, a white PSD of 1 nm rms
over a 100 Hz bandpass has been assumed, as suggested by
industrial studies \citep{VanDenTool:2006}. The output OPD PSD
indicates the total residue after correction by the FSU, limited
at low frequencies (below $\sim$ 2\,Hz) by the non-perfect control
of the input perturbations, by the noise of the FSU between 2 and
30\,Hz and by the ODL noise beyond 30\,Hz. The tip/tilt control
loop is treated in a similar way, assuming a noise of
10\,mas/$\sqrt{\rm Hz}$ per tip-tilt mirror. The same assumptions
have been considered for FKSI.

The optimised control loop performances are displayed in Table~\ref{Fig:loop_perfo}
for the GENIE instrument working on the 8-m Unit Telescopes (UT) at the VLTI (results taken from Paper\,I), the
ALADDIN instrument working on 1-m telescopes at Dome C (results taken from Paper\,II), and the space-based instruments as presented in this paper. The observations are carried out for a Sun-like G2V star located at 20\,pc on a 100 sec observation sequence using either
the 47-m UT2-UT3 baseline at the VLTI (waveband: 3.5-4.1 $\mu$m), a baseline length of 20\,m for ALADDIN
(waveband: 3.1-4.1 $\mu$m), a 40-m baseline length for \peg{} (waveband: 1.5-6.0 $\mu$m) and the 12.5-m baseline for FKSI
(waveband: 3.0-8.0 $\mu$m). As in the case of ALADDIN, dispersion and intensity errors are
expected to be very low in space and the corresponding control loops have been disabled in GENIEsim for simulating \peg{} and FKSI. This is indicated by a 0\,Hz
control loop frequency in Table~\ref{Fig:loop_perfo}. Fringe
tracking can be carried out at much lower frequencies than for
ground-based instruments (about 60\,Hz instead of 2\,kHz) and the
residual OPD errors are much lower with a typical stability of
about 2\,nm rms. Pointing errors can also be controlled at lower
frequencies ($<$ 100\,Hz instead of 1\,kHz), but the residual
tip/tilt is somewhat larger. Globally, the instrumental nulling
performance is better by at least a factor 10 with respect to
GENIE and ALADDIN because OPD errors remain the dominant
perturbations. Taking into account geometric stellar leakage, the
overall nulling performance of \peg{} is only about $10^{-3}$ due
to the combined effect of the larger baseline length and the
extension of the wavelength range towards shorter wavelengths.
Relaxing the collision avoidance requirements of 20\,m or flying
in triangular formation would enable shorter interferometric
baseline lengths and would therefore improve the overall nulling
performance of \peg{} (the geometric null is proportional to the
baseline length to the square). Another way to improve the overall
nulling performance while keeping the linear configuration is to
discard the short wavelengths. This is discussed in more details
in Section~\ref{sec:perfos}. With its 12.5-m baseline length and a
wavelength range of [3-8] $\mu$m, the total null of FKSI is about
3.0 $\times$ 10$^{-5}$. Note that the results presented for GENIE
and ALADDIN assume the ``best case scenario", which takes into
account pupil averaging, a physical phenomenon reducing the power
spectral density of piston and dispersion at high frequencies (see
paper\,I for more details).

%

\section{Simulated performance}\label{sec:perfos}



\subsection{Coupling efficiency}\label{sec:coupling}

\begin{figure}[!t]
\centering
\includegraphics[width=9.2cm]{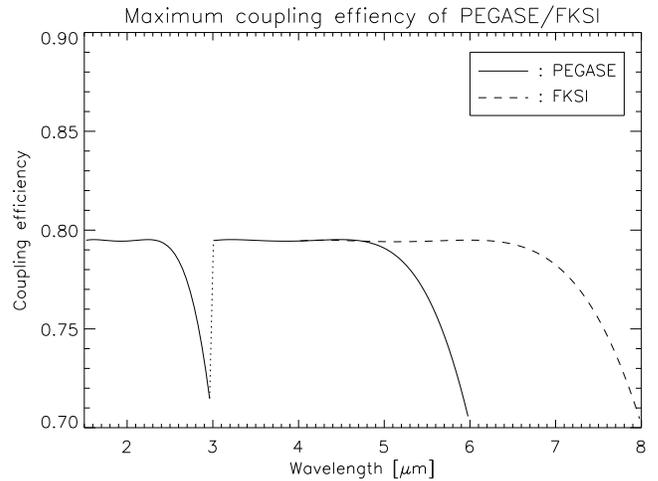}
\caption{Maximum coupling efficiency for \peg{} and FKSI with
respect to the wavelength. The core radius is chosen so as to stay
single-mode on the whole wavelength range and the focal lengths
are optimised at wavelengths of 4.5\,$\mu$m and 6\,$\mu$m,
respectively for \peg{} and FKSI. }\label{fig:coupling}
\end{figure}

The coupling efficiency represents the fraction of incoming light
from a point-like source which is transmitted into an optical
fibre. It depends on the core radius of the fibre, its numerical
aperture, the wavelength, the diameter of the telescope and its
focal length \citep{Ruilier:2001}. In order to have an efficient
correction of wave-front defects, the core radius of the fibres is
chosen so as to ensure single-mode propagation over the whole
wavelength range. The focal length can then be optimised to give
the maximum coupling efficiency at a chosen wavelength and more
importantly, to provide a roughly uniformly high coupling
efficiency across the whole wavelength band. This is generally
achieved by optimising the coupling efficiency in the middle of
the wavelength range. However, for fibres covering a wide
wavelength range, this procedure can lead to a significant
degradation of the coupling efficiency at long wavelengths (where
the instruments are most sensitive, see Section~\ref{sec:snr}).
For instance, the coupling efficiency of FKSI would be below 50\%
in the [7-8]\,$\mu$m band. For \peg, the use of two fibres partly
solves this issue but the coupling efficiency can be further
improved. Optimising the coupling efficiency at wavelengths of
4.5\,$\mu$m for \peg{} and 6\,$\mu$m for FKSI is particularly
convenient to maximize the coupling efficiency at long wavelengths
while keeping a high level at short wavelengths (see
Fig.~\ref{fig:coupling}). In both cases, the coupling efficiency
remains around its maximum (about 80\%) over almost the whole
wavelength band of each fibre and decreases to a minimum of about
70\% at the longest wavelengths. With these assumptions, we
obtained optimised focal lengths of 1.1\,m and 1.4\,m respectively
for \peg{} and FKSI. Note that in the case of \peg, we will
discard in the following study the wavelength range corresponding
to the first fibre (1.5-3.0\,$\mu$m), which is not well suited for
exozodiacal disc detection.

\subsection{Signal-to-noise ratio analysis}\label{sec:snr}

In this section, we present the different sources of noise
simulated by GENIEsim and the level at which they contribute to
the final SNR in the case of a Sun-like star located at 20\,pc.
Each source of noise is given on output of GENIEsim in
photo-electrons detected per spectral channel. Considering an
integration time of 30 min, the detailed noise budget in the
highest-SNR spectral channel is given in Table~\ref{tab:snrs} for
\peg{} and FKSI (first column). The listed sources of signal and
noise are briefly discussed hereafter.

\begin{table*}[t]
  \centering
\caption{Expected sensitivity of \peg{} (40-m baseline) and FKSI, given in number of zodis that can be detected
around a Sun-like star located at 20\,pc in 30 min. For each
instrument, the individual contributions are given in
photo-electrons in three cases: in the optimum wavelength bin, in the optimised wavelength range and in the whole wavelength range. We assume a 1\% precision on stellar diameter knowledge. The dash sign indicates that no calibration is performed.} \label{tab:snrs}
  \begin{tabular}{l | c c c| c c c}
    \hline
    \hline
      & & \peg{} & & & FKSI & \\
    \hline
    Wavelength [$\mu$m] & 5.96 & [5.7-6.0] & [3.0-6.0] & 7.83 & [6.3-8.0] & [3.0-8.0]\\
    Bandwidth [$\mu$m]  & 0.08 & 0.3 & 3.0 & 0.34 & 1.7 & 5.0 \\
    \hline
    Stellar signal [e-]           & 1.5 $\times$ $10^7$ & 6.6 $\times$ $10^7$ & 1.8 $\times$ $10^9$ & 5.0 $\times$ $10^7$ & 3.5 $\times$ $10^8$ & 3.0 $\times$ $10^9$ \\
    Raw instr. leakage [e-]       & 8.3 $\times$ $10^1$ & 3.6 $\times$ $10^2$ & 1.2 $\times$ $10^4$ & 2.7 $\times$ $10^2$ & 1.9 $\times$ $10^3$ & 2.1 $\times$ $10^4$ \\
    Total stellar leakage [e-]    & 2.2 $\times$ $10^3$ & 1.0 $\times$ $10^5$ & 6.1 $\times$ $10^5$ & 6.6 $\times$ $10^2$ & 5.4 $\times$ $10^3$ & 1.1 $\times$ $10^5$ \\
    20-zodi signal [e-]           & 1.0 $\times$ $10^3$ & 4.1 $\times$ $10^3$ & 4.9 $\times$ $10^4$ & 2.7 $\times$ $10^3$ & 1.4 $\times$ $10^4$ & 2.9 $\times$ $10^4$ \\
    Background signal [e-]        & 2.8 $\times$ $10^2$ & 7.3 $\times$ $10^2$ & 9.7 $\times$ $10^2$ & 1.8 $\times$ $10^2$ & 2.5 $\times$ $10^2$ & 2.5 $\times$ $10^2$ \\
    \hline
    Calibrated geom. leakage [e-] & 4.3 $\times$ $10^1$ & 1.9 $\times$ $10^2$ & 1.2 $\times$ $10^5$ & 8.0 $\times$ $10^0$ & 7.0 $\times$ $10^1$ & 1.8 $\times$ $10^3$ \\
    Calibrated instr. leakage [e-]& 7.2 $\times$ $10^1$ & 2.3 $\times$ $10^2$ & $- -$     & 4.4 $\times$ $10^1$ & 1.3 $\times$ $10^2$ & 1.9 $\times$ $10^3$ \\
    Instability noise [e-]        & 1.0 $\times$ $10^0$ & 4.3 $\times$ $10^0$ & 1.8 $\times$ $10^2$ & 1.4 $\times$ $10^0$ & 1.2 $\times$ $10^1$ & 1.9 $\times$ $10^2$ \\
    Shot noise [e-]               & 4.7 $\times$ $10^1$ & 1.0 $\times$ $10^2$ & 7.8 $\times$ $10^2$ & 2.6 $\times$ $10^1$ & 7.3 $\times$ $10^1$ & 3.4 $\times$ $10^2$ \\
    Detector noise [e-]           & 1.8 $\times$ $10^1$ & 3.6 $\times$ $10^1$ & 1.1 $\times$ $10^2$ & 1.8 $\times$ $10^1$ & 4.0 $\times$ $10^1$ & 8.0 $\times$ $10^1$ \\
    Background noise [e-]         & 2.7 $\times$ $10^1$ & 4.6 $\times$ $10^1$ & 1.4 $\times$ $10^2$ & 3.0 $\times$ $10^1$ & 6.3 $\times$ $10^1$ & 1.8 $\times$ $10^2$ \\
    \hline
    Zodis for SNR=5 (calibrated) & 10 & 7.8 & 34 & 2.2 & 1.3 & 9.2\\
    \hline
  \end{tabular}  
\end{table*}

\begin{itemize}
\item The stellar signal represents the total number of
photo-electrons detected in both constructive and destructive
outputs.  \item The raw instrumental leakage accounts for the
stellar photons collected at the destructive output due to the influence of instrumental
imperfections such as co-phasing errors, wavefront errors or
mismatches in the intensities of the beams. \item The 20-zodi
signal is the amount of photo-electrons at the
destructive output that come from the circumstellar disc, assumed to
be face-on and to follow the same model as in the solar system
\citep{Kelsall:1998}, except for a global density factor of 20.
\item The background signal takes into account the instrumental
brightness and the emission of the local zodiacal cloud. In the
absence of atmosphere, the latter becomes the main background
contributor and overwhelms the instrumental brightness by a factor
$\sim$1000 at 3.5\,$\mu$m or $\sim$250 at 5.5\,$\mu$m respectively
for \peg{} and FKSI. \item The geometric stellar leakage accounts
for the imperfect rejection of the stellar photons due to the
finite size of the star. Thanks to the analytical expression of
the rejection rate (see Paper\,I), it can be calibrated. Here, we
assume a typical precision ($\Delta \theta_{\star}$) of 1\% on
stellar angular diameters so that a calibration accuracy of 2\% is
reached on geometric stellar leakage. \item The raw instrumental
leakage can be decomposed into its mean value and its
variability, referred to as ``instability noise"
\citep{Lay:2004,Chazelas:2006}. The mean value can be estimated by
observing a calibrator star, provided that the interferometer
behaves in the same way during both science observation and
calibration. This calibration process is
obviously limited by its own geometrical stellar leakage,
instability noise, shot noise, detector noise and background
noise. Therefore, calibrating the mean instrumental leakage is not
necessarily useful for the improvement of the sensitivity. The
absence of calibration is indicated by a dash sign in
Table~\ref{tab:snrs}. \item Shot noise is due to the statistical
arrival process of the photons from all sources. It is mainly
dominated by stellar leakage and by the emission of the solar
zodiacal cloud. \item Detector noise is computed assuming a
read-out noise of 15 electrons rms and a typical read-out
frequency of 0.01\,Hz. \item The background noise stands for the
residual background signal in the calibration process. Two off-axis fibres located close
to the ``science" fibre in the focal plane are used to measure the background emission in real
time. Using this technique, the background
noise is reduced to the sum of the shot noise contribution from
the background itself and of the stellar light coupled into the
``background" fibres. Considering fibres located at 40\arcsec from
the axis and telescopes with a central obscuration of 14\%, the
residual stellar light in the background fibres does not exceed
about 10$^{-5}$ of the total stellar flux. More details about this
technique can be found in \cite{Absil:thesis}.
\end{itemize}

The single channel SNR can be improved by adding the signals from
different spectral channels, taking into account the possible
correlation of the noises between the wavelength bins. In this
study, we assume that systematic noises such as geometrical
leakage, instrumental leakage and instability noise are perfectly
correlated between the wavelength bins so that the noise
contributions have to be added linearly. On the other hand, random
noises such as shot noise, detector noise and background noise are
considered uncorrelated between the spectral channels and are thus
added quadratically. Combining spectral channels is efficient to a
limited extent and wide band observations give generally poor
results. This is illustrated in Table~\ref{tab:snrs} which details
the noise budget in the optimum wavelength range (second column)
and in the whole wavelength range (third column).

For both \peg{} and FKSI, the highest-SNR wavelength bin
corresponds to the longest wavelength of the science waveband with
an achievable sensitivity of respectively 10 and 2.2 zodis for a
Sun-like star located at 20\,pc. This sensitivity is slightly
improved by combining the spectral channels in the [5.7-6.0]\,$\mu$m
and [6.3-8.0]\,$\mu$m bands respectively for \peg{} and FKSI. Wider
wavelength ranges would degrade the sensitivity as illustrated by
the whole band sensitivity (respectively 34 and 9.2 zodis). This
is because the part of the SNR that is due to systematic
noises is not improved by combining spectral channels and both
\peg{} and FKSI are largely dominated by geometric stellar leakage
at short wavelengths. As a side effect, the calibration of
instrumental leakage, which is very efficient for ground-based
instruments (see Papers\,I and II), would impair the performance
of \peg{} for observations performed in the whole wavelength
range. In the optimum wavelength range of \peg, calibrating the
instrumental leakage has only a slight influence on the final
sensitivity and the geometric stellar leakage remains the dominant
noise contributor, indicating that \peg{} would present a better
sensitivity with a shorter-baseline configuration (for instance
with the three spacecraft flying in triangular formation). For
FKSI, geometric stellar leakage is less problematic due to the
shorter baseline length but remains one of the main noise
contributors. In the optimum wavelength range, the sensitivity of
FKSI is also dominated by shot noise in this particular case (a
Sun-like star located at 20\,pc). In the next section, we will see
however that geometric stellar leakage is generally dominant for
brighter targets.

\subsection{Estimated sensitivity}\label{sec:results}

Following the method used for the GENIE and ALADDIN studies (see
Papers I and II), the performances of \peg{} and FKSI are presented
for 4 hypothetic targets representative of the \darwin/TPF
catalogue \citep{Kaltenegger:2007}: a K0V star located at 5\,pc, a
G5V located at 10\,pc, a G0V located at 20\,pc and a G0V located
at 30\,pc. The results of the simulations are presented in
Table~\ref{Table:perfo}, taking into account the calibration
procedures (i.e.,~background subtraction, geometric leakage
calibration and instrumental leakage calibration) when necessary.
The detection threshold is set at a global SNR of 5 in the
optimised wavelength range. Unless specified otherwise, the
integration time has been fixed to 30 min and the accuracy on the
stellar angular diameters to 1\%.

FKSI is the most sensitive instrument and can detect circumstellar
discs with a density down to the level of the solar zodiacal
cloud. For the four representative targets of the \darwin/TPF
catalogue, FKSI can detect discs of 2.6, 1.0, 0.9 and 1.8 zodis
compared to 40, 12, 7.0 and 5.5 zodis for \peg{} (see
Table~\ref{Table:perfo}). For both instruments, geometric stellar
leakage is the dominant noise in all cases, except for the G0V
star located at 30\,pc for which FKSI is dominated by the shot and
background noises. This explains why the sensitivity decreases for
the closest targets, which have a larger angular stellar diameter
and therefore produce more geometric stellar leakage for a given
baseline length. This also explains why the optimum wavelength
range is wider for the distant targets, for which combining the
spectral channels is more efficient due to the higher relative
contribution of shot noise to the final SNR. Note also that, for
the same reason, the optimum wavelength range for a given star is
always wider for FKSI than for \peg.

The main difference between FKSI and \peg{} is related to the
geometric stellar leakage which is much larger in the
case of \peg{} due to its 40-m baseline length. Considering the
same baseline length for both instruments, the sensitivity of FKSI
would however remain better than that of \peg{} due to the longer
observing wavelength and the lower thermal background. Indeed,
observing at longer wavelengths improves the geometric stellar
rejection which is proportional to the squared wavelength. For
instance, with a hypothetic baseline length of 12.5\,m, \peg{}
could detect circumstellar discs of 10, 4.2, 3.8 and 7.7 zodis
compared to 2.6, 1.0, 0.9 and 1.8 zodis for FKSI (see
Table~\ref{Table:perfo}). The feasibility of such a flight
configuration is however beyond the scope of this paper and will
not be addressed.

\begin{figure}[t]
\centering
 \hspace{-0.5 cm}
\includegraphics[width=9.2cm]{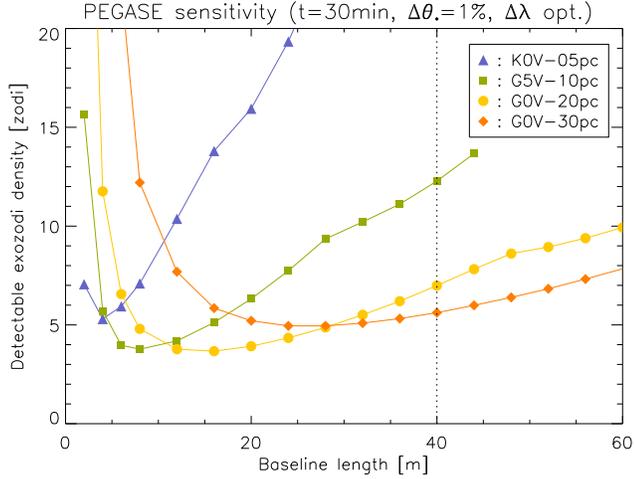}
\caption{Simulated performance of \peg{} for four typical
\darwin/TPF targets with respect to the baseline length, assuming
1\% uncertainty on stellar angular diameters and an integration
time of 30 min.} \label{fig:zodi_vs_base}
\end{figure}

\begin{figure}[!t]
\centering \hspace{-0.5 cm}
\includegraphics[width=9.2cm]{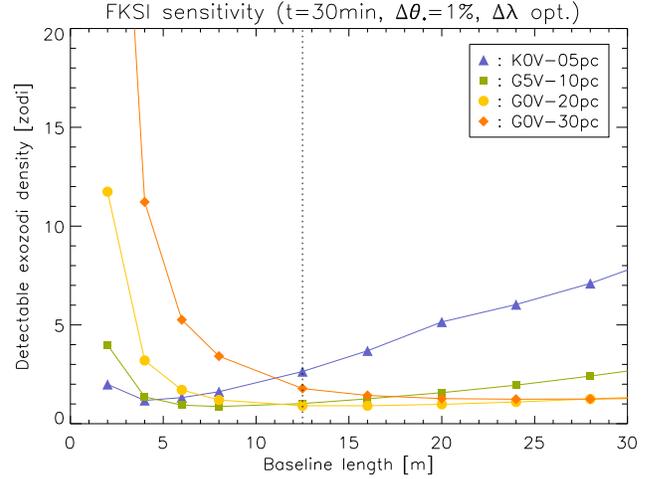}
\caption{Simulated performance of FKSI for four typical
\darwin/TPF targets with respect to the baseline length, assuming
1\% uncertainty on the stellar angular diameter and an integration
time of 30 min.} \label{fig:zodi_vs_base_FKSI}
\end{figure}

The estimated sensitivity is
represented as a function of baseline length in Fig.~\ref{fig:zodi_vs_base} for \peg{} and in
Fig.~\ref{fig:zodi_vs_base_FKSI} for FKSI, where the wavelength
range is optimised separately for each baseline length. As already
suggested, the sensitivity at long baseline lengths is dominated
by geometric stellar leakage, especially for the closest targets
which have a larger stellar angular diameter. By reducing the
baseline length, the starlight rejection improves and the
sensitivity curves decrease towards a minimum, indicating the
optimum baseline length. It is interesting to note that the 12.5-m
interferometric baseline of FKSI is a good compromise for most stars in the \darwin/TPF catalogue. The
decrease in performance towards longer baselines lengths is
stronger for \peg{} than for FKSI since it observes at shorter
wavelengths. At short baseline lengths, background noise becomes
dominant due to the decrease
of the exozodiacal disc transmission and the sensitivity curves rise again. The slight inflection in the
sensitivity curves of Fig.~\ref{fig:zodi_vs_base} (e.g., at a
baseline length of 15\,m for the K0V star) indicates the baseline
lengths at which the instrumental leakage calibration becomes
useless and would not improve the sensitivity. The difference in
sensitivity between \peg{} and FKSI decreases with the target
distance since the optimum baseline length of \peg{} is getting
closer to 40\,m.

\begin{table}[!t]
  \setlength{\tabcolsep}{4.2pt}
  \centering
    \caption{Simulated sensitivity and optimum wavelength range of \peg{} and FKSI for four representative targets of the \darwin/TPF catalogue,
  assuming 1\% uncertainty on the stellar angular diameter and an integration time of 30 min.}\label{Table:perfo}
  \begin{tabular}{c|c|c|c}
    \hline
    \hline
               & \peg\,-\,12.5\,m & \peg\,-\,40\,m & FKSI \\
    \begin{tabular}{c}
     Targets \\
    \end{tabular}&
    \begin{tabular}{c c}
     zodi & $\lambda$ [$\mu$m]\\
    \end{tabular}&
    \begin{tabular}{c c}
     zodi & $\lambda$ [$\mu$m]\\
    \end{tabular}&
    \begin{tabular}{c c}
     zodi & $\lambda$ [$\mu$m]\\
    \end{tabular}\\
    \hline
    \begin{tabular}{c}
    K0V\,-\,05\,pc\\
    G5V\,-\,10\,pc\\
    G0V\,-\,20\,pc\\
    G0V\,-\,30\,pc\\
    \end{tabular} &
    \begin{tabular}{c c}
    10 & 5.9\,-\,6.0 \\
    4.2 & 5.6\,-\,6.0 \\
    3.8 & 5.0\,-\,6.0 \\
    7.7 & 4.3\,-\,6.0 \\
    \end{tabular} &
    \begin{tabular}{c c}
    40 & 5.9\,-\,6.0 \\
    12 & 5.9\,-\,6.0 \\
    7.0 & 5.7\,-\,6.0 \\
    5.5 & 5.4\,-\,6.0 \\
    \end{tabular} &
    \begin{tabular}{c c}
    2.6 & 7.6\,-\,8.0\\
    1.0 & 7.2\,-\,8.0\\
    0.9 & 6.7\,-\,8.0\\
    1.8 & 6.0\,-\,8.0\\
    \end{tabular}\\
    \hline
    \end{tabular}
\end{table}

%

\section{Discussion}\label{sec:discussions}
\subsection{Comparison with ground-based sites}

\begin{figure}[t]
\centering
\includegraphics[width=9.2cm]{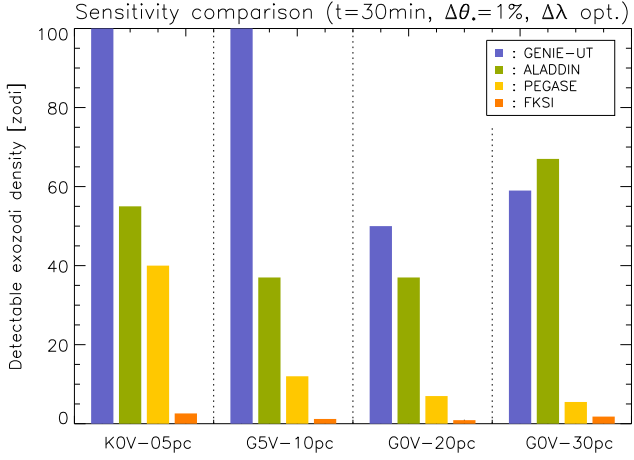}
\caption{Expected performance for \peg{} and FKSI compared to the
ground-based instruments (for 30 min integration time and 1\%
uncertainty on the stellar angular diameters).} \label{fig:tri}
\end{figure}

In order to provide a fair comparison between ground- and
space-based nulling interferometers, we use the performance
estimations of GENIE and ALADDIN obtained with GENIEsim for the 4
representative targets of the \darwin/TPF catalogue (from
Papers\,I and II). The detectable exozodiacal dust densities for GENIE
on the unit telescopes (UT - 8\,m diameter), ALADDIN, \peg{} and
FKSI are represented in Fig.~\ref{fig:tri}, considering an
integration time of 30\,min and an uncertainty on the stellar
angular diameters of 1\%. For all target stars, the space-based
nulling interferometers are the most sensitive instruments. \peg{}
(resp.\,FKSI) outperforms ALADDIN by a factor ranging from about
1.5 (resp.\,20) for the K0V star located at 5\,pc to a factor of
about 10 (resp.\,30) for the G0V star located at 30\,pc. This
better sensitivity of space-based instruments is mainly due to the
lower thermal background and geometric stellar leakage, which are
the dominant noises for GENIE and ALADDIN. While the absence of
atmosphere in space and the cooler optics explain the lower
thermal background, the longer observing wavelength improves the
geometric stellar rejection, which is proportional to the squared
wavelength.

As discussed in the previous sections, \peg{} and FKSI are
generally limited by geometric stellar leakage. Reducing the
baseline length to improve the sensitivity is not possible either
due to the free-flying constraints for \peg{} or due to the fixed
boom on FKSI. Besides reducing the interferometric baseline
length, another way to minimize the geometric stellar leakage is
to improve the knowledge on stellar angular diameters. Considering
the four targets representative of the \darwin/TPF catalogue,
Table~\ref{tab:results} gives the sensitivity to exozodiacal discs
of \peg{} and FKSI for different uncertainties on the stellar
angular diameter. The results of GENIE on the unit telescopes (UT
- 8\,m diameter) and ALADDIN are also presented for comparison.
Unlike the other instruments, FKSI is relatively insensitive to
the uncertainty on the stellar angular diameter, with a
sensitivity below 4 zodis even for a knowledge of the stellar
angular diameter of 1.5\%.


\begin{table}[t]
\centering
\caption{Performance comparison between GENIE, ALADDIN, \peg{} and
FKSI expressed in detectable exozodiacal disc densities as
compared to the solar zodiacal disc (for different uncertainties
on the stellar angular diameter and an integration time of 30
min).} \label{tab:results}
\begin{tabular}{c|c c c c|c}
  \hline
  \hline
  Star & 0.25\% & 0.5\% & 1\% & 1.5\% & Instrument \\
  \hline
            & 110 & 230 & 450 & 680 & GENIE - UT \\
  K0V - 05pc & 20 & 33 & 55 & 79 &  ALADDIN \\
            & 12 & 21 & 40 & 60 & PEGASE \\
            & 0.9 & 1.4 & 2.6 & 3.9 & FKSI \\
  \hline
             & 30 & 59 & 120 & 180 & GENIE - UT \\
  G5V - 10pc & 15 & 24 & 37 & 51 &  ALADDIN \\
             & 4.7 & 8.3 & 12 & 17 & PEGASE \\
             & 0.5 & 0.7 & 1.0 & 1.4 & FKSI \\
  \hline
             & 21 & 29 & 50 & 73 & GENIE - UT \\
  G0V - 20pc & 19 & 25 & 37 & 48 & ALADDIN \\
             & 2.8 & 4.2 & 7.0 & 9.5 & PEGASE \\
             & 0.7 & 0.8 & 0.9 & 1.1 & FKSI \\
  \hline
             & 36 & 46 & 59 & 71 & GENIE - UT \\
  G0V - 30pc & 62 & 63 & 67 & 72 &  ALADDIN \\
             & 3.1 & 3.9 & 5.5 & 7.3 & PEGASE \\
             & 1.7 & 1.7 & 1.8 & 1.9 & FKSI \\
    \hline
\end{tabular}
\end{table}

\subsection{Influence of integration time}

\begin{figure}[t]
\centering
\includegraphics[width=9.2cm]{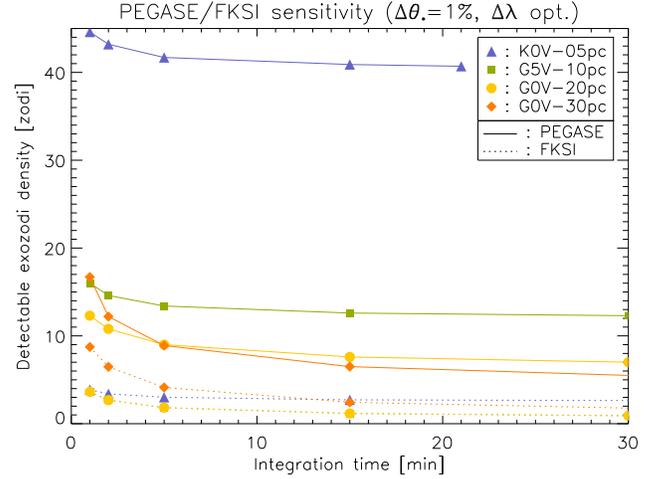}
\caption{Simulated performance of \peg{} and FKSI in terms of
exozodiacal disc detection with respect to the integration time,
for an uncertainty on the stellar angular diameter of 1\% and in
the optimised wavelength range.}\label{Fig:zodi_vs_intt}
\end{figure}

Increasing the integration time has different influences on the
individual noise sources. For instance, shot noise, detector noise
and instability noise (to the first order) have the classical
t$^{1/2}$ dependance and their relative impact on the final SNR
decreases for longer integration times. On the other hand, the
imperfect calibration of geometric and instrumental stellar
leakage is proportional to time, so that increasing the
integration time has no influence on the associated SNR. Since
geometric stellar leakage is generally dominant, increasing the
integration time does not improve significantly the sensitivity to
exozodiacal discs. The sensitivity as a function of the
integration time is represented in Fig.~\ref{Fig:zodi_vs_intt},
using the optimum wavelength range. With a 40-m baseline length,
\peg{} is dominated by geometric stellar leakage for the four
targets and reducing the integration time to five minutes has
almost no influence. For FKSI, geometric stellar leakage is not
dominant for the G0V star located at 30\,pc and increasing the
integration time improves slightly the sensitivity (1.4-zodi disc
detectable in 60 minutes instead of 1.8-zodi disc in 30 min). For
the other three targets, geometric stellar leakage is dominant and
integration times longer than 30 minutes have no significative
influence on the sensitivity. Like \peg, an integration time of
five minutes is already sufficient to reach the maximum
sensitivity for most targets. For comparison, ALADDIN could reach a
sensitivity of 30 zodis after about 8 hours of integration time
for G0V stars located between 20 and 30\,pc (see Paper\,II). Due to the low thermal background,
\peg{} and FKSI achieve their maximum sensitivity much faster than
ground-based nulling instruments.

\subsection{Influence of telescope diameter}

\begin{figure}[t]
\centering
\includegraphics[width=9.2cm]{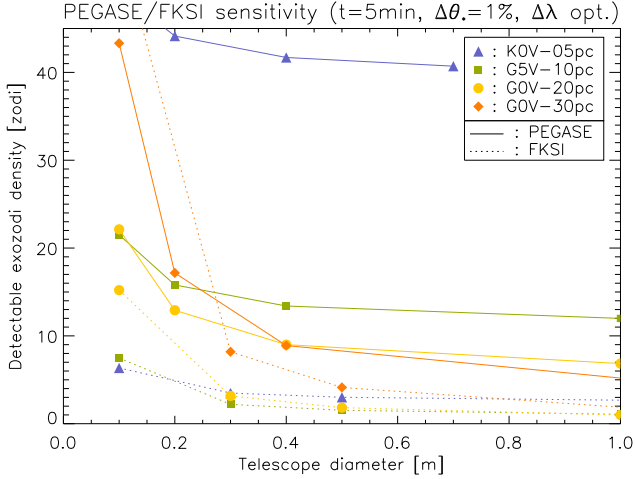}
\caption{Simulated performance of \peg{} and FKSI in terms of
exozodiacal disc detection for different pupil diameters,
considering an uncertainty on the stellar angular diameter of 1\%
($\Delta \theta_{\star}$) and 5 min of
integration.}\label{Fig:zodi_vs_diam}
\end{figure}

Similarly to integration time, increasing the telescope diameter
has different influences on the individual noise sources. Since
the geometric nulling ratio does not depend on the aperture size,
the component of the SNR which is due to geometric stellar leakage
is not improved by increasing the telescope diameter. Since the
geometric stellar leakage is generally dominant for an integration
time of 30\,min, different pupil sizes have therefore little
influence on the final sensitivity. In order to clearly show
the impact of different pupil sizes, we consider in this section
an integration time of 5\,min which is generally sufficient to
reach the maximum sensitivity (see previous section). Considering
an uncertainty on the stellar angular diameter of 1\% and 5\,min
of integration time, the sensitivities of \peg{} (solid lines) and
FKSI (dashed lines) for different pupil diameters are presented in
Fig.~\ref{Fig:zodi_vs_diam}. As expected, the sensitivity varies
more significantly for the faintest targets, which are more
dominated by shot noise. For \peg, the sensitivity is already
close to the maximum with the 40-cm diameter apertures and
increasing the telescope diameter has only a slight impact for the
faintest target (G0V star located at 30\,pc). For FKSI, the
sensitivity remains practically unchanged for telescopes with a
diameter larger than 30\,cm, except for the G0V star located at
30\,pc. In practice, the final choice of the pupil diameter will
result from a trade-off between integration time, feasibility and
performance.

\subsection{Sky coverage}

In the context of \darwin/TPF preparatory activities, another
relevant issue is the sky coverage, i.e.,\ the part of the
celestial sphere accessible by each instrument. A
representative way to assess the sky coverage is to determine how
many stars of the \darwin/TPF all sky target catalogue
\citep{Kaltenegger:2007} can be observed by each instrument. This
value depends on the combination of two parameters: the location
of the instrument and its pointing direction ability. For the
ground-based instruments, we assume that the zenith distance can
not be larger than 60\degre. For space-based instruments, the
pointing direction covers the part of the sky with an
ecliptic latitude between $\pm$ 30\degre\,for \peg{} and $\pm$
20\degre\,for FKSI (after 1 year of observation).

\begin{figure}[t]
\centering
\includegraphics[width=9.2cm]{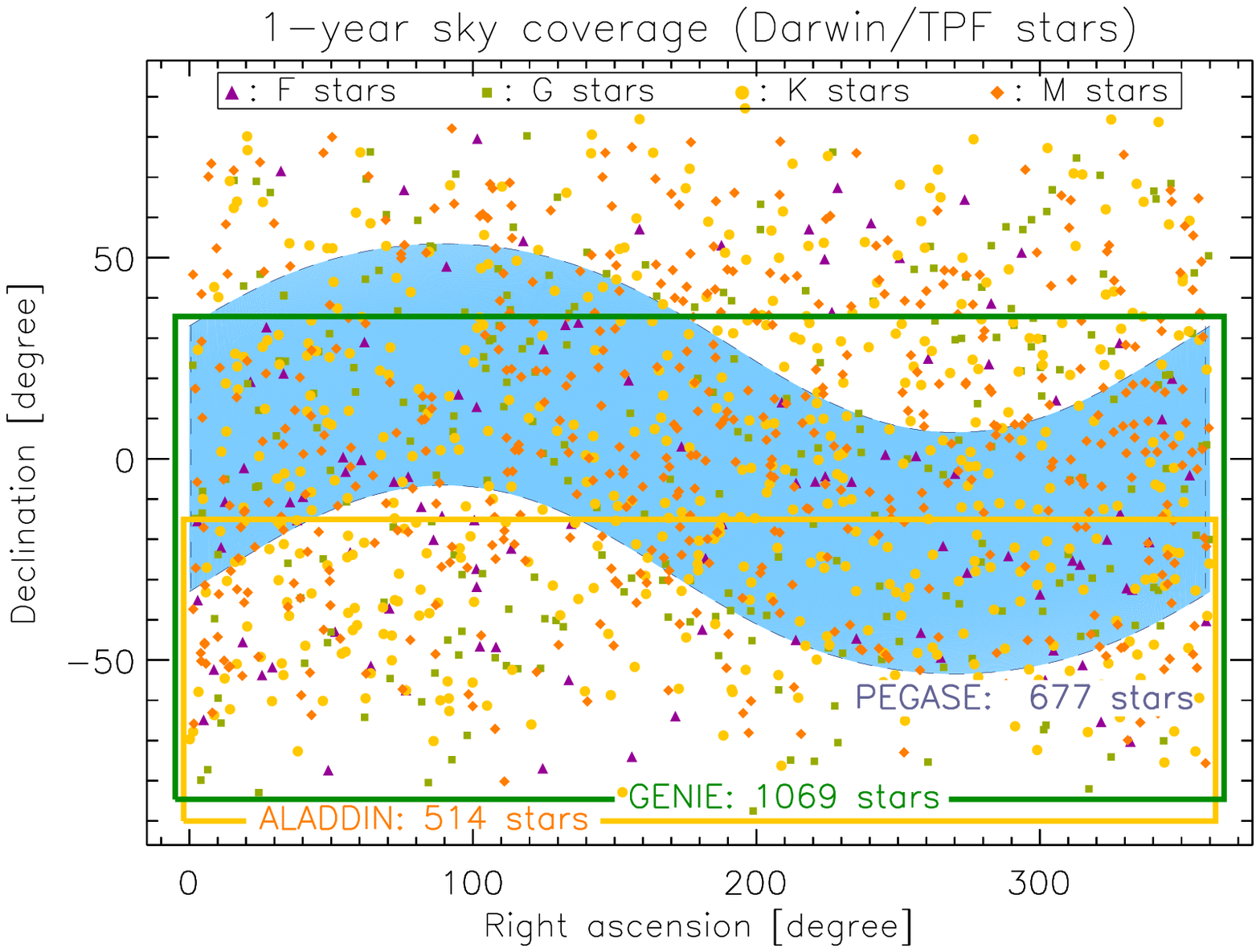}
\caption{Sky coverage after 1 year of observation of GENIE (dark
frame), ALADDIN (light frame) and \peg{} (shaded area) shown with
the \darwin/TPF all sky target catalogue. The blue-shaded area
shows the sky coverage of a space-based instrument with an
ecliptic latitude in the [-30\degre,30\degre] range (such as
\peg). The sky coverage of FKSI is similar to that of \peg{} with
an extension of 40\degre\,instead of
60\degre.}\label{Fig:skycov}
\end{figure}

Considering the 1354 single target stars of the \darwin/TPF
catalogue (106 F, 251 G, 497 K and 500 M stars), the results are
presented in Fig.~\ref{Fig:skycov} for GENIE (dark frame), ALADDIN
(light frame) and \peg{} (shaded area). The sky coverage of FKSI
is not represented for the sake of clarity but is similar to that
of \peg{} with an extension in declination of 40\degre\,instead
of 60\degre. The stars enclosed in a specified frame are
observable by the corresponding instrument in a 1-year observation
window. Counting the stars in each frame, GENIE can observed 1069
targets (90 F, 191 G, 405 K, 383 M stars), ALADDIN 514 (52 F, 98
G, 204 K, 160 M stars), \peg{} 677 (53 F, 125 G, 244 K and 255 M
stars) and FKSI 443 (28 F, 74 G, 164 K and 177 M stars). These
values correspond to about 80\%, 40\%, 50\% and 30\% of the
targets, respectively for GENIE, ALADDIN, \peg{} and FKSI. Note that ALADDIN
and the space-based instruments cover complementary regions of the
sky and are able to survey most of the targets with a declination
lower than 50\degre.

%

\section{Conclusions}

Nulling interferometry is a promising technique to assess the
level of circumstellar dust in the habitable zone around nearby
main sequence stars. From the ground, instruments like GENIE (VLTI
nuller, using two 8-m telescopes) and ALADDIN (Antarctic nuller,
using two dedicated 1-m telescopes) could achieve the detection of
exozodiacal discs with a density of several tens of zodis.
The high Antarctic plateau is a particularly well suited site in that context, so that ALADDIN is expected to achieve
the best sensitivity (down to 30 zodis in few hours of integration time).
Observing from space provides the solution to go beyond this sensitivity by getting rid
of the high thermal background constraining ground-based
observations. In this paper, we have investigated the
performance of two space-based nulling interferometers which have
been intensively studied during the past few years (namely \peg{}
and FKSI). Even though they have been initially designed for the
characterisation of hot extrasolar giant planets, \peg{} and FKSI
would be very efficient to probe the inner region of circumstellar
discs where terrestrial habitable planets are supposed to be
located. Within a few minutes, \peg{} (resp.\,FKSI) could detect
exozodiacal discs around nearby main sequence stars down to a
density level of 5 (resp.\,1) times our solar zodiacal cloud and thereby
outperform any ground-based instrument. FKSI can achieve this
sensitivity for most targets of the \darwin/TPF catalogue while
\peg{} becomes less sensitive for the closest targets with
detectable density levels of about 40 times the solar zodiacal
cloud. This outstanding and uniform sensitivity of FKSI over the
\darwin/TPF catalogue is a direct consequence of the short
baseline length (12.5\,m) used in combination with an appropriate
observing wavelength of about 8\,$\mu$m, which is ideal for
exozodiacal disc detection. Another advantage of FKSI is to be
relatively insensitive to the uncertainty on stellar angular
diameters, which is a crucial parameter driving the performance of
other nulling interferometers. In terms of sky coverage, we show that these
space-based instruments are able to survey about 50\% of the \darwin/TPF target stars.
The sky coverage reaches 80\% if they are used in combination with
ALADDIN, which provides a complementary sky coverage. 
Beyond the technical demonstration of nulling
interferometry in space, the present study indicates that \peg{}
and FKSI would be ideal instruments to prepare future life-finding
space missions such as \darwin/TPF.

\begin{acknowledgements}
The authors are grateful to Lisa Kaltenegger (Harvard-Smithsonian
Center for Astrophysics) for providing the updated \darwin{}
catalogue, Jean Surdej (IAGL), Arnaud Magette (IAGL), Pierre Riaud
(Paris observatory), Tupper Hyde (NASA/GSFC), Pierre-Yves Guidotti
(CNES) and Julien Morand (EADS-Astrium) for their support and
contribution. The authors also thank Richard Barry for a careful
reading of the manuscript. The first author acknowledges the
support of the Belgian National Science Foundation (``FRIA").
O.A.\ acknowledges the financial support from the European
Commission's Sixth Framework Program as a Marie Curie
Intra-European Fellow (EIF).
\end{acknowledgements}

\bibliography{biblio}
\bibliographystyle{aa}

\end{document}